\documentclass{article}
\usepackage{ijcai26}

\usepackage{times}
\usepackage{soul}
\usepackage{url}
\usepackage[hidelinks]{hyperref}
\usepackage[utf8]{inputenc}
\usepackage[small]{caption}
\usepackage{graphicx}
\usepackage{amsmath}
\usepackage{amsthm}
\usepackage{booktabs}
\usepackage{algorithm}
\usepackage{algorithmic}
\usepackage[switch]{lineno}
\usepackage{multirow}
\usepackage[capitalize]{cleveref}
\usepackage{mathtools}
\usepackage{amssymb}

\usepackage{graphicx}
\makeatletter
\newcommand{\figcaption}{\def\@captype{figure}\caption}
\makeatother

\usepackage{amsmath}

\usepackage{amssymb}

\crefname{section}{Sec.}{Secs.}
\Crefname{section}{Section}{Sections}
\Crefname{table}{Table}{Tables}
\crefname{table}{Tab.}{Tabs.}

\title{MusicWeaver: Programmable Long-Form Music Generation with
Provably Local Editing}

\author{
Xuanchen Wang
\and
Heng Wang\and
Weidong Cai\\
\affiliations
The University of Sydney\\
\emails
xwan0579@uni.sydney.edu.au,
\{heng.wang, tom.cai\}@sydney.edu.au
}

\begin{document}

\maketitle

\begin{figure*}[t]
    \centering
    \includegraphics[width=\textwidth]{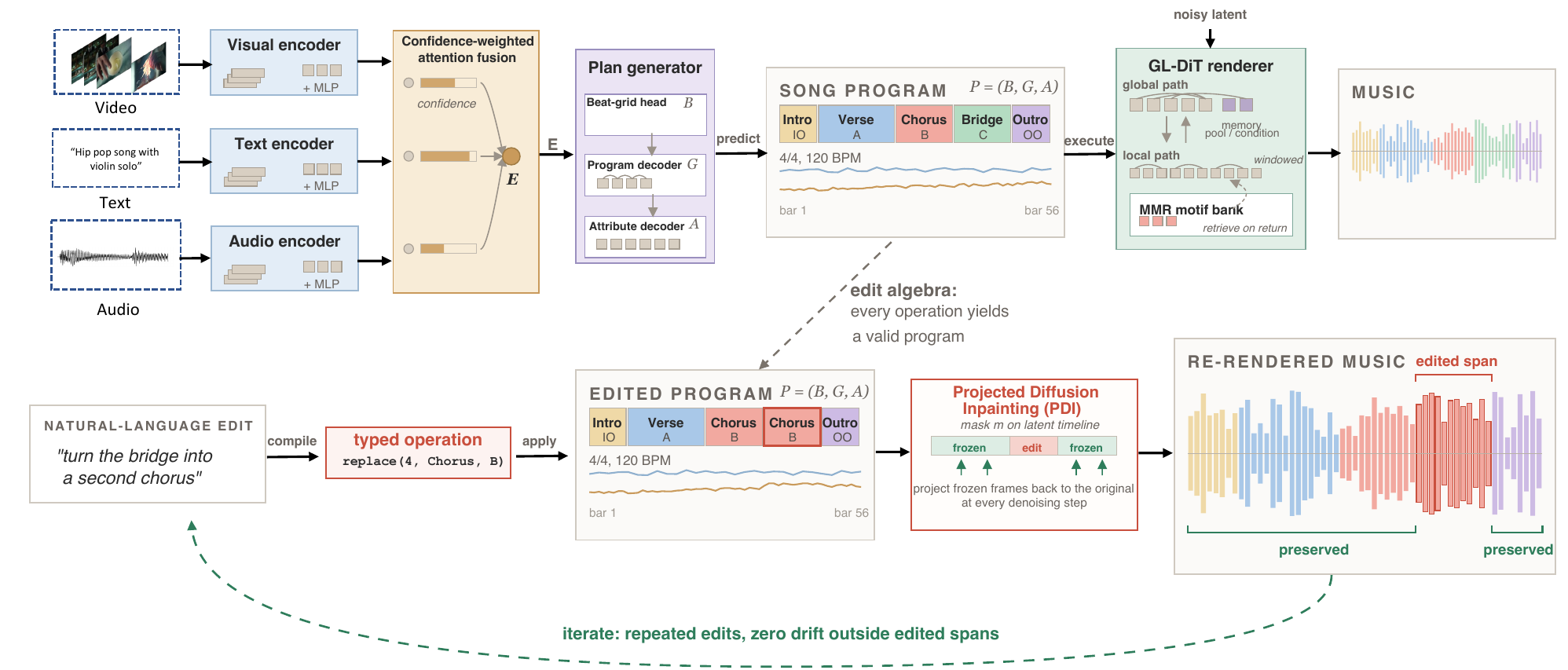}
    \caption{
    Overview of MusicWeaver. Top: prompts from any combination of
    text, video, and audio are encoded, fused by confidence-weighted
    attention, and compiled by the plan generator into a human-interpretable
    song program; alternatively, a program is induced from an existing
    recording. The GL-DiT renderer with motif memory retrieval executes the
    program into minute-scale audio. Bottom: a natural-language instruction
    is compiled into a typed operation of the edit algebra, which always
    yields a valid program, and Projected Diffusion Inpainting re-renders
    only the affected span while audio outside it is preserved exactly
    across repeated revisions.
    }
    \label{fig:overview}
\end{figure*}

\begin{abstract}
Music generation systems produce increasingly realistic audio, yet they expose no
interface between a creator's structural intent and the rendered sound. Form can
only be steered through prompt engineering, and local revisions require
regenerating entire pieces. We recast music creation as program-guided generation, where an explicit, human-interpretable program layer serves as the intermediate representation between intent and audio. We present MusicWeaver, which instantiates this idea by decomposing music generation into planning and rendering. The planning stage predicts a structured plan, a multi-level song program that encodes musical form, motif recurrence, and bar-level musical attributes. The rendering stage then synthesizes audio conditioned on this plan. We formalize editing as an algebra of typed operations over plans, covering section replacement, insertion, deletion, motif retagging, and attribute changes, and we prove that these operations preserve plan validity. To realize edits, we propose Projected Diffusion Inpainting, which guarantees by construction that audio outside the edited span is preserved exactly, no matter how many revision rounds are applied. For rendering, we design a Global-Local Diffusion Transformer with Motif Memory Retrieval that executes plans at minute scale and produces section returns that are consistent yet varied. Beyond generation, our framework supports plan induction, which
recovers an editable program from an existing recording, and we build a
natural-language editor that compiles free-form instructions into validated
operation sequences. We further introduce plan-faithfulness and edit-fidelity
measures and validate them against human judgments. Experiments on generation
conditioned on text, video, and their combination show that our method achieves
state-of-the-art structural coherence and editability, and its edit quality
remains stable over sequential revisions where prior methods degrade. 
\end{abstract}


\section{Introduction}

Music generation has advanced rapidly, with autoregressive and diffusion-based
systems producing audio of increasingly convincing quality from text
prompts~\cite{copet2023musicgen,liu2023audioldm,liu2024audioldm2,evans2025stableaudioopen,ziv2024masked}
and, more recently, sustaining generation over full-song
durations~\cite{evans2024longform,yuan2025yue,ning2025diffrhythm,gong2025ace}.
Yet these advances have not translated into practical creation tools. The reason
is not audio quality but the absence of an interface. A creator who wants to
replace a verse with a chorus, insert a bridge, or reshape the energy arc of a
piece has no direct way to express these intentions. Form can only be steered
through prompt engineering, whose effect on structure is indirect and
unreliable, and any local revision requires regenerating the entire piece,
discarding material the creator wished to keep.

This gap is fundamental rather than incidental. Real composition is iterative.
Creators revise a small set of bars many times while expecting the rest of the
piece to remain untouched, and they reason about music through structural
abstractions such as sections, motifs, and arrangement rather than through
waveforms or latent codes. Supporting this workflow imposes two requirements
that current systems do not meet jointly. First, structural intent must be
expressible and editable at the level composers think in. Second, edits and
generation alike must remain coherent at minute scale, with stable tempo,
convincing motif returns, and clear sectional form. Long-context
generators~\cite{copet2023musicgen,evans2024longform,yuan2025yue} extend
duration but expose no explicit handle on form, while instruction-driven
editing methods~\cite{wang2023audit,han2024instructme,zhang2024musicmagus}
operate directly on audio with only weak locality, so intended local edits leak
into surrounding regions and repeated edits accumulate drift.

We argue that both requirements can be met by introducing an explicit,
human-interpretable program layer between intent and audio, and we recast music
creation as \emph{program-guided generation}. In this view, a musical piece is
described by a program that a human can read and modify, and audio synthesis
becomes the execution of that program. An explicit musical program turns structural edits into discrete, semantically meaningful operations with precisely controlled scope, rather than opaque perturbations of a prompt or latent representation.

We present MusicWeaver, which instantiates program-guided generation by decomposing creation into planning and rendering (Figure~\ref{fig:overview}). The planning stage predicts a structured plan, a multi-level song program encoding a beat grid, a
segment-level form with explicit motif recurrence, and bar-level attributes
such as harmony, groove, energy, and density. The rendering stage synthesizes
audio conditioned on this plan. To make the plan a genuine interface rather
than a passive annotation, we formalize editing as an algebra of typed
operations over plans, covering section replacement, insertion, deletion, motif
retagging, and attribute changes, and we show that every operation is guaranteed to produce a well-formed, renderable plan, so no sequence of edits can leave the program in a broken state. Edits are realized in audio by Projected Diffusion
Inpainting (PDI), a sampling mechanism that projects all content outside the
edited span back to its original latent values at every denoising step. This
yields a locality guarantee that holds by construction rather than by training:
unedited audio is preserved exactly, over arbitrarily many revision rounds.

Executing programs at minute scale places demands on the renderer that standard
diffusion transformers~\cite{peebles2023dit} do not meet, since full attention
over minutes of latent frames is computationally prohibitive while blockwise
generation dilutes long-range identity. We therefore design a Global-Local
Diffusion Transformer (GL-DiT) that couples a compressed global pathway,
which tracks bar-level progression with persistent memory, with a local pathway
that synthesizes fine acoustic detail. A Motif Memory Retrieval (MMR) module
stores compact representations of realized motifs and retrieves them when the
program calls for a return, so recurring sections preserve their identity with
controllable variation instead of drifting or degenerating into copies.

A program layer is only useful if programs are easy to obtain and easy to
modify. Our framework therefore supports plan induction, which recovers an
editable program from an existing recording, extending composer-style editing
to real music rather than only to our own generations. We further build a
natural-language editor that compiles free-form instructions into sequences of
validated operations from the edit algebra, so that users can revise a piece
conversationally while retaining every formal guarantee of the underlying
operations. Finally, because established audio metrics do not measure whether
a system follows a program or realizes an edit, we introduce plan-faithfulness
and edit-fidelity measures and validate them against human judgments.

Our contributions are summarized as follows:
\begin{itemize}
    \item We recast music creation as program-guided generation and propose
    MusicWeaver, a two-stage framework in which a human-interpretable,
    multi-level song program serves as the intermediate representation between
    creative intent and rendered audio.
    \item We formalize composer-style editing as an algebra of typed plan
    operations that always yield well-formed, renderable plans, and we propose
    Projected Diffusion Inpainting, which guarantees by construction that
    content outside the edited span is preserved exactly across repeated
    revisions.
    \item We design the Global-Local Diffusion Transformer with Motif Memory
    Retrieval to execute programs at minute scale, combining long-range
    structural memory with high-resolution synthesis and consistent yet varied
    motif returns.
    \item We broaden the interface and its evaluation through plan induction
    from existing recordings, a natural-language editor that compiles
    instructions into validated operations, and new plan-faithfulness and
    edit-fidelity measures validated against human judgments.
\end{itemize}

\section{Related Work}
\paragraph{Music and Audio Generation.}
Deep generative models have advanced from text-to-audio
synthesis~\cite{liu2023audioldm,ghosal2023tango,majumder2024tango2,liao2024baton}
to high-quality text-to-music
generation~\cite{copet2023musicgen,liu2024audioldm2,evans2025stableaudioopen,ziv2024masked}
and full-song generation with long-window diffusion or foundation
models~\cite{evans2024longform,yuan2025yue,gong2025ace,ning2025diffrhythm}.
Video-to-music and multimodal systems further condition music on visual or
cross-modal cues~\cite{di2021cmt,kang2024video2music,li2024diffbgm,lin2025vmas,tian2025vidmuse,tian2025audiox}.
These methods scale duration and modality, but musical form remains implicit:
users cannot directly inspect, verify, or revise section structure and motif
recurrence. MusicWeaver addresses this missing interface by placing an editable
program between intent and audio.

\paragraph{Controllable and Intermediate Representations.}
Prior work introduces explicit controls such as chords, beats, tempo, key,
melody, dynamics, and rhythm~\cite{melechovsky2024mustango,wu2024musiccontrolnet}.
Symbolic representations such as MIDI and lead sheets also provide
composer-facing structure, and two-stage systems generate symbolic music before
synthesis~\cite{wang2024wholesong,lu2023musecoco}. However, existing controls
are usually local, while symbolic pipelines cover only score-representable
music and often rely on fixed synthesis. MusicWeaver instead uses a coarser
audio-domain program that captures form, recurrence, and bar-level attributes,
making it human-editable, inducible from recordings, and renderable by a
diffusion model.

\paragraph{Audio and Music Editing.}
Instruction-driven methods edit audio or latent representations using text
guidance~\cite{liu2023audioldm,wang2023audit,han2024instructme,zhang2024musicmagus}. Recent work further improves source consistency through
score-distillation-based editing~\cite{niu2026steermusic} and attention-guided
structure preservation~\cite{yang2026melodia}, while symbolic systems support
localized, human-in-the-loop music infilling~\cite{hu2025compose}. However,
audio-domain methods still specify edit scope implicitly, so changes may leak
outside the intended region and repeated revisions can accumulate drift.
MusicWeaver instead compiles edits into typed operations over an explicit
program and realizes them with PDI, making locality a construction-time
guarantee rather than an empirical tendency.

\begin{figure}[t]
    \centering
    \includegraphics[width=\linewidth]{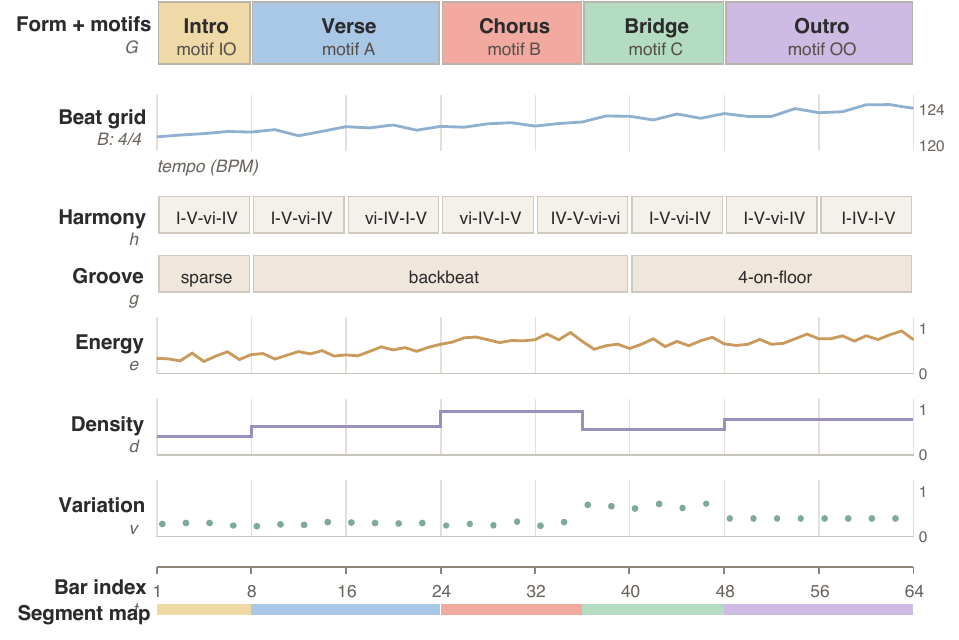}
    \caption{
    An example song program and one edit applied to it. The
    program describes a 64-bar piece through the form track $\mathcal{G}$
    with explicit motif identifiers, the beat grid $\mathcal{B}$, and the
    bar-level attribute track $\mathcal{A}$, with the segment map $\kappa(t)$
    shown under the bar axis.
    }
    \label{fig:program}
\end{figure}

\begin{figure*}[t]
    \centering
    \includegraphics[width=\textwidth]{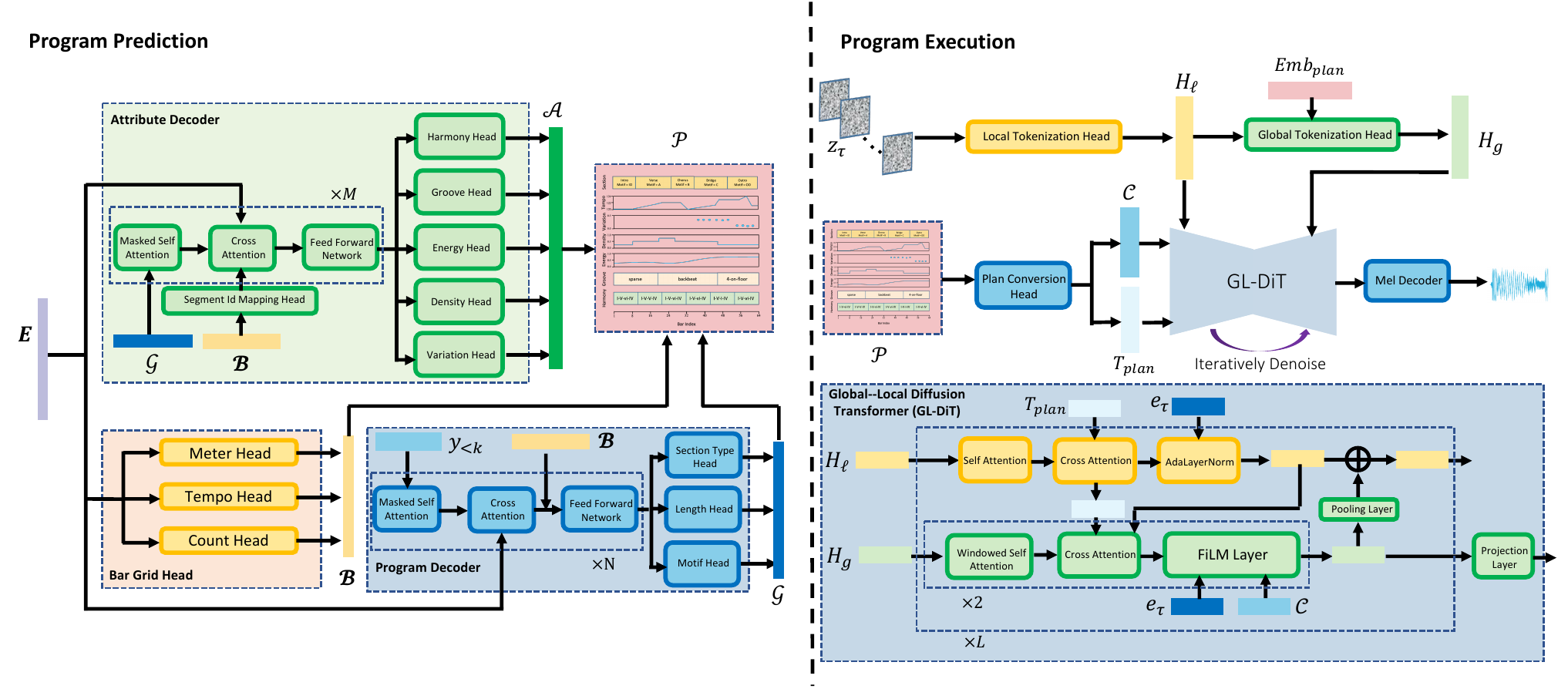}
    \caption{Architecture of MusicWeaver. Left: program prediction. The
    fused prompt embedding $E$ conditions a beat-grid head, an
    autoregressive program decoder that generates the form track
    $\mathcal{G}$ with masked self-attention and cross-attention to $E$,
    and an attribute decoder that expands the program into the bar-level
    track $\mathcal{A}$ through the segment map. Right: program
    execution. The program is converted into time-aligned controls
    $\mathcal{C}$ and program tokens $T_{\mathrm{plan}}$, and the GL-DiT
    denoiser couples a global pathway over bar tokens and persistent
    memory with a local pathway over latent patches, using windowed
    self-attention, cross-attention to global and program tokens, and
    FiLM-injected controls, to iteratively denoise the mel-VAE latent.}
    \label{fig:arch}
\end{figure*}

\section{Method}
Figure~\ref{fig:overview} summarizes the framework: prompts are fused and
compiled into a song program (Sections~\ref{sec:program}
and~\ref{sec:prediction}), the program is executed by the GL-DiT renderer
(Section~\ref{sec:renderer}), and edits are typed operations realized by
PDI (Sections~\ref{sec:algebra} and~\ref{sec:pdi}). Full proofs,
architectural details, and implementation recipes are in Appendix.

\subsection{The Song Program}
\label{sec:program}
MusicWeaver represents a piece by an explicit, human-interpretable
program; generation, editing, and induction are all defined with respect
to this single representation. A program is a triple:
\begin{equation}
    \mathcal{P} = (\mathcal{B}, \mathcal{G}, \mathcal{A}).
\end{equation}
The beat grid $\mathcal{B} = (m, T_b, \boldsymbol{\omega})$ fixes meter,
bar count, and a per-bar tempo curve, anchoring all structure to musical
time so that edits are specified in bars and mapped deterministically to
audio spans. The form track
$\mathcal{G} = \{(s_k, \ell_k, r_k)\}_{k=1}^{K}$ describes $K$ contiguous
segments with section type $s_k$, length $\ell_k$ in bars, and motif id
$r_k$; segments tile the piece, inducing a segment map
$\kappa : \{1, \dots, T_b\} \to \{1, \dots, K\}$. Shared motif ids
declare that segments realize the same material, which the renderer is
obligated to honor; the value \textsc{new} introduces fresh material.
The attribute track $\mathcal{A} = \{a_t\}_{t=1}^{T_b}$ assigns each bar
a tuple $a_t = (h_t, g_t, e_t, d_t, v_t)$ of harmony, groove, energy,
density, and variation strength, deliberately coarser than notes so the
program stays readable while the renderer supplies detail.

We call a program \emph{valid} when (W1) segment lengths tile the piece,
$\sum_k \ell_k = T_b$ with $\ell_k \geq 1$; (W2) every bar carries
attributes within their domains; (W3) every motif id other than
\textsc{new} refers to a motif introduced by an earlier segment; and
(W4) the tempo curve is total and bounded. Validity is the contract
between the program and the rest of the system: a valid program is
renderable, and its bars map unambiguously to audio through $\mathcal{B}$
and $\kappa$. Figure~\ref{fig:program} shows an example program and one
edit applied to it.

\subsection{The Edit Algebra}
\label{sec:algebra}
We define editing as an algebra of typed operations rather than free-form
modification. The algebra contains five operations:
\textsc{replace}$(k, s', r')$ rewrites segment $k$'s type and motif,
keeping its length; \textsc{insert}$(k, s, \ell, r)$ creates a segment
before position $k$, growing the bar count and extending the tempo curve
by boundary interpolation; \textsc{delete}$(k)$ removes segment $k$ with
its bars; \textsc{retag}$(k, r')$ changes only a motif id; and
\textsc{setattr}$(t_1, t_2, f, x)$ assigns an in-domain value to one
attribute field on a bar span. Rewritten or created bars receive
attributes from the motif source for returns and from type-conditional
defaults otherwise, and remain editable. Admissibility is minimal:
indices exist, lengths are positive, values are in range. Motif
references need no guard because introduction is positional: the earliest
segment carrying an id is its introduction, so recurrence stays well
defined under any reordering or removal. If $\mathcal{P}$ is valid and $o$ is an admissible operation, then $o(\mathcal{P})$ is
valid; hence any finite composition of admissible operations preserves
validity. The proof checks W1 to W4 per operation and is given in Appendix. 

Each operation has a bounded \emph{footprint} $\sigma(o)$, the set of
bars whose content it may change: the target segment for
\textsc{replace} and \textsc{retag}, the inserted bars for
\textsc{insert}, the span for \textsc{setattr}, and the junction bars for
\textsc{delete}. Through the beat grid, a footprint maps
deterministically to a span of latent frames; in
Section~\ref{sec:pdi} the PDI mask is derived from $\sigma(o)$, so the
locality the algebra specifies symbolically is exactly the locality the
renderer guarantees acoustically. Richer edits arise by composition, and
operation sequences are deterministic, reproducible, and form the compile
target of the natural-language editor (Section~\ref{sec:nleditor}).

\subsection{Program Prediction}
\label{sec:prediction}
Programs are predicted from prompts or induced from recordings
(Section~\ref{sec:nleditor}); we describe prediction here. Text, video,
and audio prompts are processed by frozen pretrained encoders, projected
into a shared space by learned MLPs, and combined by a
confidence-weighted attention module into a fused embedding $E$, which
keeps prediction agnostic to which modalities were supplied. A program is
then predicted coarse to fine:
\begin{equation}
    p(\mathcal{P} \mid E) =
    p(\mathcal{B} \mid E)\,
    p(\mathcal{G} \mid \mathcal{B}, E)\,
    p(\mathcal{A} \mid \mathcal{G}, \mathcal{B}, E).
    \label{eq:factorization}
\end{equation}
Categorical heads on $E$ predict meter and bar count, and tempo is
parameterized by knots interpolated to a per-bar curve; both may instead
be user-specified. The form track is generated autoregressively by a
Transformer decoder attending to $E$, with segment type, length, and
motif predicted by factorized heads. Lengths exceeding the remaining bar
budget are masked during decoding, enforcing W1 at generation time, and
motif ids are predicted over previously introduced motifs plus
\textsc{new}, so W3 holds by construction: the predictor emits only valid
programs, and the guarantees the algebra provides for editing hold for
generation as well. A second decoder expands the program into per-bar
attributes conditioned on $E$, the beat grid, and the segment context
$(s_{\kappa(t)}, r_{\kappa(t)})$. The three predictors and their conditioning are shown in Figure~\ref{fig:arch} (left).

The predictor is trained by maximum likelihood with teacher forcing on
prompt and reference-program pairs, where reference programs are produced
by the induction pipeline of Section~\ref{sec:nleditor} over the training
corpus, so the label format is identical to the programs users edit.
Prediction quality is therefore upper-bounded by induction quality; we
quantify both in the experiments.

\subsection{Program Execution}
\label{sec:renderer}
Executing a program over minutes demands tracking section-level
progression, for which full attention over the latent timeline is
prohibitive, while synthesizing fine acoustic detail, which compressed
long-context modeling alone cannot provide. Rendering operates as conditional latent diffusion in the space of a mel-VAE~\cite{rombach2022ldm,liu2023audioldm} with a standard noising schedule~\cite{ho2020ddpm} and DDIM sampling~\cite{song2021ddim}. The program conditions the denoiser through two streams: a time-aligned control
sequence $\mathcal{C}$ carrying per-bar fields resampled to the latent
timeline through the beat grid, and a short program token sequence
$T_{\mathrm{plan}}$ summarizing structure and recurrence, injected by
cross-attention.

\paragraph{Global-Local Diffusion Transformer.}
The denoiser is a Global-Local Diffusion Transformer (GL-DiT), shown in
Figure~\ref{fig:arch} (right). At each diffusion step $\tau$, it takes the noisy
latent $z_\tau$, timestep embedding $e_\tau$, time-aligned controls
$\mathcal{C}$, and program tokens $T_{\mathrm{plan}}$, and predicts
$\hat{\epsilon}$.

We first split $z_\tau$ into non-overlapping $(p_t,p_f)$ patches and project
them into local tokens:
\begin{equation}
    H_\ell^{(0)} = W_p\,\mathrm{Patch}(z_\tau) + P,
\end{equation}
where $P$ is a 2D positional embedding and $\mathcal{C}$ is pooled onto the
same patch grid. To track long-range program state, we also form global tokens
at bar resolution:
\begin{equation}
    H_g = \big[\,\mathrm{PoolBar}(H_\ell) + \mathrm{Emb}_{\mathrm{plan}}
    \;;\; H_{\mathrm{mem}}\,\big],
\end{equation}
where $\mathrm{PoolBar}(H_\ell)$ produces one token per bar,
$\mathrm{Emb}_{\mathrm{plan}}$ encodes bar-aligned plan fields, and
$H_{\mathrm{mem}}$ are learned persistent memory tokens.

GL-DiT stacks $L$ coupled blocks~\cite{peebles2023dit}. Each block updates
global tokens with full self-attention and cross-attention to
$T_{\mathrm{plan}}$, modulated by $e_\tau$ through AdaLayerNorm. Local tokens
are then updated by windowed self-attention, cross-attention to global and
program tokens, and FiLM-injected controls~\cite{perez2018film}. Local features
are pooled back into bar summaries after each block, grounding the global state
in the synthesized audio. After $L$ blocks, the local tokens are projected back
to latent space to predict $\hat{\epsilon}$.

\paragraph{Motif Memory Retrieval.}
A program's recurrence declarations are obligations: segments sharing a
motif id must sound like realizations of the same material. Long-context
modeling alone dilutes this identity when attention windows are limited,
producing returns that drift or collapse into copies. Motif Memory
Retrieval (MMR) enforces the obligation explicitly at render time. MMR
maintains a memory bank mapping each motif id $r$ to motif tokens $V_r$.
After rendering a segment with id $r$, the local tokens within the
segment span $\Omega_k$ are pooled and projected into motif tokens by a
lightweight projector; if $V_r$ exists, it is updated by an exponential
moving average, allowing controlled evolution across occurrences. When
rendering a segment declared as a return of $r$, the stored $V_r$ is
retrieved and injected as additional cross-attention context in every
local layer, so each patch attends to the realized identity of the motif
while synthesizing the current segment. The program's variation field
$v_t$ gates the strength of this conditioning, so the same mechanism
spans faithful reprises and loose recollections, under the composer's
control rather than the sampler's.



\subsection{Guaranteed-Local Edit Rendering}
\label{sec:pdi}
Projected Diffusion Inpainting (PDI) builds on per-step replacement
sampling from diffusion image inpainting~\cite{lugmayr2022repaint,song2021score}, making locality a property of the sampler rather than a
tendency of the model; PDI differs in deriving its mask from program
footprints, stating the preservation guarantee exactly for the dilated
mask, and extending replacement to length-changing edits via splicing. Given an
operation or operation sequence, its footprint maps through the beat grid
to a binary mask $m \in \{0, 1\}^{T' \times 1}$ on the latent timeline;
to avoid audible seams, the editable region is dilated by a short
transition band, and the guarantee below is stated for the dilated mask,
so every unmarked frame is untouched. With
$z_0^{\mathrm{old}}$ the latent of the current render, or of real audio
when editing an induced program, sampling starts from:
\begin{equation}
    z_T = (1 - m) \odot z_0^{\mathrm{old}} + m \odot \epsilon,
    \qquad \epsilon \sim \mathcal{N}(0, I),
\end{equation}
and each DDIM proposal $\tilde{z}_{\tau - 1}$ under the edited program's
conditioning is projected back onto the constraint set:
\begin{equation}
    z_{\tau - 1} = m \odot \tilde{z}_{\tau - 1}
    + (1 - m) \odot z_0^{\mathrm{old}}.
    \label{eq:projection}
\end{equation}

For every step $\tau$ and frame $i$ with $m_i = 0$,
$z_\tau[i] = z_0^{\mathrm{old}}[i]$; hence over any finite sequence of
edit sessions, frames outside the union of masks are identical to the
original latent, independent of the number of sessions. The proof is immediate from Eq.~\ref{eq:projection} by induction. This is
the formal content of drift-free iteration: preservation is enforced per
step, not encouraged by training. The guarantee is stated in latent
space; the mel-VAE decoder and vocoder have a fixed receptive-field
horizon across mask boundaries, beyond which we verify waveforms are
sample-identical, as reported in the experiments. Operations that change
the bar count are handled by splicing frozen latent spans to their new
positions on the edited timeline and marking the inserted region or
junction band editable, so copied content keeps its identity while moving
in time; details are in Appendix. Finally, masked training episodes
apply the diffusion loss under the projection constraint with random
masks and optional in-mask plan edits, improving boundary continuity
without weakening the guarantee, which the sampler enforces regardless.

\begin{table}[t]
\centering\small
\resizebox{\linewidth}{!}{
\begin{tabular}{l l c c c c c}
\toprule
Method & Task & KL$\downarrow$ & FD$\downarrow$ & FAD$\downarrow$ &
Align.$\uparrow$ & SCS$\uparrow$ \\
\midrule
AudioLDM-2 & T2M
& 1.28 & 16.68 & 2.92 & 0.22 & 63.2 \\
Stable Audio Open & T2M
& 1.49 & 34.52 & 3.15 & 0.22 & 71.2 \\
MAGNET & T2M
& 1.35 & 21.26 & 4.54 & 0.19 & 75.5 \\
AudioX & T2M
& \underline{1.05} & \textbf{11.23} & 1.73 & \underline{0.23} & 76.8 \\
ACE-Step & T2M
& 1.07 & 11.64 & \underline{1.62} & \textbf{0.24} & \underline{80.3} \\
DiffRhythm & T2M
& 1.18 & 14.92 & 1.88 & 0.21 & 77.9 \\
MusicWeaver & T2M
& \textbf{1.02} & \underline{11.58} & \textbf{1.51} & \underline{0.23} & \textbf{83.2} \\
\midrule
VidMuse & V2M
& 0.78 & 28.31 & 2.42 & 0.21 & 75.4 \\
AudioX & V2M
& \underline{0.69} & \textbf{23.25} & \underline{2.23} & \underline{0.23} & \underline{78.8} \\
MusicWeaver & V2M
& \textbf{0.66} & \underline{24.38} & \textbf{2.12} & \textbf{0.24} & \textbf{81.5} \\
\midrule
AudioX & TV2M
& \underline{0.49} & \underline{19.54} & \underline{1.49} & \underline{0.22} & \underline{78.8} \\
MusicWeaver & TV2M
& \textbf{0.45} & \textbf{17.67} & \textbf{1.45} & \textbf{0.24} & \textbf{80.9} \\
\bottomrule
\end{tabular}}
\caption{Generation quality and structural coherence. Best and second-best
results within each task group are shown in \textbf{bold} and
\underline{underline}, respectively. Align. is measured by CLAP for
text-conditioned generation and ImageBind-AV for video-conditioned generation.}
\label{tab:main}
\end{table}

\begin{table}[t]
\centering\small
\resizebox{\linewidth}{!}{
\begin{tabular}{l c c c c c c}
\toprule
Method & \multicolumn{2}{c}{60 s} & \multicolumn{2}{c}{120 s}
& \multicolumn{2}{c}{180 s} \\
\cmidrule(lr){2-3}\cmidrule(lr){4-5}\cmidrule(lr){6-7}
& SCS$\uparrow$ & FAD$\downarrow$ & SCS$\uparrow$ & FAD$\downarrow$
& SCS$\uparrow$ & FAD$\downarrow$ \\
\midrule
Stable Audio long-form
& 73.4 & 2.28 & 68.2 & 2.77 & 61.5 & 3.34 \\
DiffRhythm
& 76.9 & 1.98 & 69.8 & 2.51 & 63.1 & 3.08 \\
ACE-Step
& 80.3 & 1.69 & 75.6 & 2.05 & 70.2 & 2.57 \\
MusicWeaver
& \textbf{84.1} & \textbf{1.55}
& \textbf{82.6} & \textbf{1.76}
& \textbf{80.8} & \textbf{2.06} \\
\bottomrule
\end{tabular}}
\caption{Minute-scale coherence across different generation durations.}
\label{tab:duration-data}
\end{table}

\subsection{Plan Induction and the Natural-Language Editor}
\label{sec:nleditor}
Induction recovers a program from arbitrary audio in the same coarse to
fine order as prediction: a neural beat and downbeat tracker~\cite{bock2016madmom} yields the bar grid and tempo, structure segmentation over learned audio embeddings~\cite{li2024mert} yields section boundaries snapped to downbeats with types assigned by a segment-level classifier, motif ids are induced by clustering segment
embeddings with the earliest occurrence as introducer, matching the
positional convention of Section~\ref{sec:algebra}, and bar attributes
are estimated per bar from chord recognition, onset patterns, loudness,
onset counts, and embedding distance to the motif's introduction. The
result is valid by construction. Induced programs label the training
corpus (Section~\ref{sec:prediction}), admit composer-style editing of
real recordings, and define the reference for plan faithfulness in the
experiments; pipeline components and accuracy are detailed in Appendix.

The natural-language editor compiles free-form instructions into
operation sequences. A language model receives the serialized program,
the instruction, and the operation signatures, and is constrained to emit
operation calls in a structured schema with no free-form mutation
available. A validator replays admissibility against the program state,
returns violations for one round of repair, and rejects failing
sequences outright, so every guarantee of Sections~\ref{sec:algebra} and~\ref{sec:pdi} holds regardless of model behavior: the model proposes, the algebra disposes. The compiled sequence is shown to the user as a program diff before rendering.

\section{Experiments and Results}
\label{sec:experiments}
\subsection{Setup}

\noindent\textbf{Datasets.}
We train MusicWeaver on VGGSound-Caps and V2M-Caps~\cite{tian2025audiox,chen2020vggsound} using the official preprocessing and splits. The combined training set contains about 220K clips,
corresponding to 610 hours of audio after filtering. Structured programs are
obtained using the induction pipeline in Section~\ref{sec:nleditor}. We evaluate text-to-music generation on MusicCaps~\cite{agostinelli2023musiclm}, video-conditioned generation on V2M-bench~\cite{tian2025vidmuse}, and minute-scale generation on MW-Long, a held-out set with 60, 120, and 180 second generations.

\noindent\textbf{Implementation.}
GL-DiT uses $L{=}12$ global-local blocks, width $d{=}768$, 12 heads, and
$K_m{=}16$ memory tokens per block. Audio is resampled to 24 kHz and
represented with 128-bin mel spectrograms. We train with
$T_{\mathrm{diff}}{=}1000$ diffusion steps and sample using 50-step DDIM.
The final model is trained for 600K updates on 8 NVIDIA A100-80GB GPUs.
Unless otherwise specified, results are averaged over three seeds.

\noindent\textbf{Baselines.}
For text-to-music, we compare with
AudioLDM-2~\cite{liu2024audioldm2}, Stable Audio
Open~\cite{evans2025stableaudioopen}, MAGNET~\cite{ziv2024masked},
AudioX~\cite{tian2025audiox}, ACE-Step~\cite{gong2025ace}, and
DiffRhythm~\cite{ning2025diffrhythm}; for minute-scale generation we
additionally compare with Stable Audio long-form~\cite{evans2024longform}.
For video-conditioned generation, we compare with
VidMuse~\cite{tian2025vidmuse} and AudioX. For editing, we compare with
AUDIT~\cite{wang2023audit}, InstructME~\cite{han2024instructme}, and
MusicMagus~\cite{zhang2024musicmagus}. We use official checkpoints
whenever available and evaluate all systems with the same duration,
loudness normalization, and metric implementations.

\noindent\textbf{Standard Metrics.}
We report KL divergence and Fr\'echet Distance (FD) computed with
PANNs~\cite{kong2020panns}, and Fr\'echet Audio Distance
(FAD)~\cite{kilgour2019fad} for audio fidelity. Conditional alignment
is measured by CLAP~\cite{wu2023clap} for text and
ImageBind-AV~\cite{girdhar2023imagebind} for video. For editing, we
report Chord Recognition Accuracy (CRA), Pitch Class Histogram
similarity (PCH), and Inter-Onset Interval consistency (IOI),
following~\cite{han2024instructme}.

\noindent\textbf{Proposed Evaluation Metrics.}
Standard metrics do not evaluate our two target capabilities, so we
introduce the Structure Coherence Score (SCS), aggregating five
audio-only structural sub-scores; the Edit Fidelity Score (EFS),
combining edit realization inside the target region with the fraction
of change concentrated there; and Plan Faithfulness (PF), the agreement
between a program and the plan re-induced from its render, at the form,
motif, and attribute levels. Formal definitions, weights, and the
freeze-before-evaluation protocol are in the appendix. To validate the
metrics, 20 participants with music production or MIR experience rated
generated and edited clips; SCS correlates most strongly with human
structure ratings and EFS with edit correctness and locality
(details in Appendix).

\subsection{Result Analysis }
\noindent\textbf{Generation Quality and Structure.}
Table~\ref{tab:main} shows that MusicWeaver achieves the strongest overall trade-off:
best or second-best on every fidelity and alignment metric across T2M,
V2M, and TV2M, and the highest SCS in every task group. The consistent
SCS margin over baselines that match us on fidelity suggests the gains
come from program conditioning and motif-aware rendering rather than
from raw synthesis capacity.


\noindent\textbf{Minute-Scale Coherence.}
All methods degrade with duration (Table~\ref{tab:duration-data}), but MusicWeaver degrades
least, losing 3.3 SCS points from 60 to 180 seconds against 10 or more
for every baseline, indicating that explicit programs, global memory,
and motif retrieval preserve section identity precisely where
long-window attention alone begins to fail.


\noindent\textbf{Edit Quality and Locality.}
On a 300-instruction benchmark spanning seven edit categories,
MusicWeaver leads on all metrics (Table~\ref{tab:edit}), with the largest gain on
EFS: plan-space editing with localized re-rendering realizes the
intended change while preserving the rest, whereas latent-space editors
without a bar-level plan trade one for the other.


\begin{table}[t]
\centering\small
\begin{tabular}{l c c c c}
\toprule
Method & CRA$\uparrow$ & PCH$\uparrow$ & IOI$\uparrow$ & EFS$\uparrow$ \\
\midrule
AUDIT
& 73.2 & 62.8 & 59.4 & 68.8 \\
InstructME
& 74.1 & 64.5 & 60.2 & 67.8 \\
MusicMagus
& 75.6 & 65.1 & 61.7 & 70.4 \\
MusicWeaver
& \textbf{78.8} & \textbf{66.4} & \textbf{65.8} & \textbf{73.8} \\
\bottomrule
\end{tabular}
\caption{Editing quality on the instruction-based music editing benchmark.}
\label{tab:edit}
\end{table}

\begin{table}[t]
\centering\small
\resizebox{\linewidth}{!}{
\begin{tabular}{l c c c c}
\toprule
Method & $k{=}1$ & $k{=}2$ & $k{=}4$ & $k{=}8$ \\
\midrule
AUDIT & 0.037 & 0.075 & 0.136 & 0.211 \\
InstructME & 0.034 & 0.068 & 0.126 & 0.196 \\
MusicMagus & 0.029 & 0.056 & 0.108 & 0.171 \\
MusicWeaver & \textbf{0.004} & \textbf{0.007} & \textbf{0.011} & \textbf{0.014} \\
MusicWeaver, outside decoder horizon
& \textbf{0.000} & \textbf{0.000} & \textbf{0.000} & \textbf{0.000} \\
\bottomrule
\end{tabular}}
\caption{Outside-region spectral distance under repeated editing. Lower is
better.}
\label{tab:drift}
\end{table}

\noindent\textbf{Drift Under Iteration.}
Table~\ref{tab:drift} measures spectral distance outside the edited regions over $k$
sequential edits. Baselines accumulate drift roughly linearly in $k$,
while MusicWeaver is unchanged outside a decoder horizon of 16 latent
frames (about 0.68 s) and exactly zero beyond it. This demonstrates that PDI is particularly important for iterative composer-style workflows, where users revise a small region multiple times while expecting the rest of the piece to remain stable.


\noindent\textbf{Natural-Language Editing.}
We evaluate the plan editor on 300 natural-language edit instructions. The
compiler succeeds on 88.7\% of instructions without repair and 96.0\% with one
automatic repair round. Invalid edit sequences, such as negative section
length, overlapping frozen and editable spans, or motif references to
non-existent sections, are rejected in all tested cases. The most common
failure cases involve underspecified temporal references, such as ``make the
second part more energetic'', where the intended section boundary is ambiguous.

\noindent\textbf{Induction Accuracy.}
We evaluate the program induction pipeline on a manually annotated subset of
600 clips. The beat tracker obtains 94.2 beat F1, and the section boundary
detector obtains 78.1 boundary F1. Section type classification reaches 74.6\%
accuracy, motif-link prediction reaches 70.2 F1, chord recognition reaches
68.4\% accuracy, and groove classification reaches 73.9\% accuracy. These
values indicate that the induced programs are imperfect but reliable enough to
provide training supervision. To reduce the impact of noisy labels, all plan
losses are weighted by induction confidence.

\noindent\textbf{Plan Faithfulness.}
Table~\ref{tab:pf} reports PF for each plan attribute. MusicWeaver obtains the
highest overall PF because the renderer receives both time-aligned control
streams and global program tokens. Text-conditioned baselines can sometimes
match genre or energy descriptions, but they struggle with exact section
layout, motif recurrence, and bar-level harmonic or groove control.

\begin{table}[t]
\centering\small
\begin{tabular}{l c c c c}
\toprule
Method & Form & Motif & Attr. & PF$\uparrow$ \\
\midrule
AudioLDM-2
& 48.6 & 42.3 & 55.8 & 51.8 \\
AudioX
& 56.4 & 49.1 & 63.7 & 58.7 \\
ACE-Step
& 61.2 & 55.6 & 68.4 & 62.9 \\
MusicWeaver
& \textbf{86.1} & \textbf{80.5} & \textbf{85.7} & \textbf{84.7} \\
\bottomrule
\end{tabular}
\caption{Plan faithfulness on held-out programs. Form measures section layout
and boundary timing, Motif measures recurrence identity, and Attr. measures
bar-level harmony, groove, energy, density, and variation controls.}
\label{tab:pf}
\end{table}

\begin{table}[t]
\centering\small
\begin{tabular}{l c c c c}
\toprule
Variant & SCS$\uparrow$ & EFS$\uparrow$ & FAD$\downarrow$ & PF$\uparrow$ \\
\midrule
MusicWeaver (full)
& \textbf{83.2} & \textbf{72.6} & \textbf{1.51} & \textbf{84.7} \\
w/o confidence weighting
& 79.1 & 69.6 & 1.59 & 80.2 \\
w/o global path
& 74.6 & 70.9 & 1.58 & 76.1 \\
w/o local path
& 70.2 & 66.8 & 1.85 & 74.8 \\
w/o MMR
& 77.0 & 70.8 & 1.55 & 80.5 \\
MMR w/o variation gating
& 80.4 & 71.4 & 1.53 & 82.3 \\
w/o PDI, soft mask
& 83.0 & 63.5 & 1.63 & 84.1 \\
PDI, hard mask, no band
& 83.0 & 69.2 & 1.56 & 84.3 \\
w/o masked training
& 82.7 & 67.2 & 1.60 & 83.8 \\
\bottomrule
\end{tabular}
\caption{Ablations on MusicCaps.}
\label{tab:ablation}
\end{table}

\noindent\textbf{Ablations.}
Table~\ref{tab:ablation} confirms the division of labor: the global path carries
structure (largest SCS drop when removed), the local path carries
fidelity and edit quality, MMR and its variation gating carry
recurrence, and PDI with masked training carries editability, with
confidence weighting mattering most when modality cues conflict.


\noindent\textbf{User Study.}
We conduct two user studies for generation and editing. In the perception
study, 20 participants rate anonymized outputs on overall quality (OVL),
prompt relevance (REL), and coherence (COH) using 30 prompts from MusicCaps
and 30 from V2M-bench. MusicWeaver obtains the highest average scores on both
T2M (OVL 82.6, REL 85.1, COH 84.3) and V2M (OVL 72.4, REL 75.0, COH 78.6).
In the revision-workflow study, participants edit an initial clip using either
MusicWeaver or an instruction-based baseline. MusicWeaver reduces regeneration
attempts from 3.4 to 1.8 and improves perceived edit controllability from 63.5
to 81.2, indicating that the explicit program layer makes iterative revision
more predictable.

\section{Conclusion}
We recast music creation as program-guided generation and presented
MusicWeaver, in which a human-interpretable song program mediates
between intent and audio, executed at minute scale by a Global-Local
diffusion transformer with motif memory retrieval. Two guarantees
bracket the creation loop: every edit operation yields a valid program, and PDI preserves all audio outside an edit's footprint exactly, over arbitrarily many revisions. Plan induction and a natural-language editor extend the interface to real recordings and conversational use. Experiments show that MusicWeaver improves generation quality, structural coherence, plan faithfulness, and localized editing over strong baselines.

\clearpage

\bibliographystyle{named}
\bibliography{ijcai26}

\clearpage

\appendix

\section{Method}

\subsection{The Edit Algebra}
\label{sec:algebra}

\paragraph{Validity Preservation}
\label{prop:closure}
Let $\mathcal{P}$ be a valid program and $o$ an admissible operation. Then
$o(\mathcal{P})$ is valid. Consequently, for any finite sequence of
admissible operations $o_n \circ \dots \circ o_1$, the resulting program
$(o_n \circ \dots \circ o_1)(\mathcal{P})$ is valid. We check conditions W1 to W4 per operation. \textsc{replace}, \textsc{retag}, and \textsc{setattr} leave $K$, all $\ell_k$, and $T_b$ unchanged, so W1 and W4 are untouched; \textsc{setattr} preserves W2 by its domain guard, and \textsc{replace} and \textsc{retag} preserve W3 because
motif introduction is positional. \textsc{insert} and \textsc{delete}
change $T_b$ and the segment list consistently, adding or removing exactly
the bars of one segment, so segment lengths again tile the piece (W1);
attributes and tempo are defined on precisely the surviving or created bars
by construction (W2, W4); and positional introduction transfers to the
earliest remaining occurrence of each id (W3). Composition follows by
induction. The full proof is in the appendix.

\paragraph{Edit Footprint.}
Each operation touches a bounded region of the piece. We define the
footprint $\sigma(o) \subseteq \{1, \dots, T_b'\}$ as the set of bars of
the resulting program whose content the operation may change:
the bars of segment $k$ for \textsc{replace} and \textsc{retag}, the
inserted bars for \textsc{insert}, the span $[t_1, t_2]$ for
\textsc{setattr}, and for \textsc{delete} the empty set apart from the
junction bars adjacent to the removed segment. Through the beat grid, a
footprint in bars maps deterministically to a span of latent frames. This
mapping is the contract between the algebra and rendering: in
Section~\ref{sec:pdi}, the PDI mask is derived from $\sigma(o)$, so the
locality that the algebra specifies symbolically is exactly the locality
the renderer guarantees acoustically. For a sequence of operations applied
before a single re-render, the mask is the union of their footprints.


\subsection{Program Prediction}
\label{sec:prediction}




\paragraph{Beat-Grid Head.}
Meter $m$ and bar count $T_b$ are predicted by categorical heads on $E$.
The tempo curve is parameterized compactly as $K_\omega$ knots
$\tilde{\boldsymbol{\omega}} \in \mathbb{R}^{K_\omega}$, interpolated to
the per-bar curve $\boldsymbol{\omega} \in \mathbb{R}^{T_b}$. Each head is
a two-layer MLP. Meter and bar count may also be user-specified, in which
case the head outputs are overridden and the remaining factors condition
on the given values.

\paragraph{Program Decoder.}
The form track $\mathcal{G}$ is generated autoregressively over segments.
At step $k$, a Transformer decoder consumes the previously generated
segment tokens $y_{<k}$, embedded with positional and beat-grid control
embeddings, and attends to $E$ as external memory, producing a hidden
state $h_k$. Lengths that exceed the budget are masked during decoding, which enforces the tiling condition W1 at generation time rather than by rejection. Motif ids are predicted over the vocabulary of previously introduced motifs extended with a
\textsc{new} token, so W3 holds by construction: a return can only name
material that exists. The predictor therefore emits only valid programs,
and the same guarantees the algebra provides for editing
(Section~\ref{sec:algebra}) hold for generation.

\paragraph{Attribute Decoder.}
A second Transformer decoder expands the global program into per-bar
attributes. For each bar $t$, the decoder conditions on $E$, the beat grid,
and the segment context $(s_{\kappa(t)}, r_{\kappa(t)})$ through the
segment map, and multi-head categorical layers predict the attribute tuple
$a_t$. Conditioning on the motif id lets bars of recurring segments
receive attributes consistent with earlier realizations of the same
material, while the variation field $v_t$ remains free to modulate how
strictly the renderer should enforce that consistency.

\subsection{Program Execution}
\label{sec:renderer}


\paragraph{Latent Diffusion Setup.}
Rendering operates in the latent space of a mel-VAE. Let
$M \in \mathbb{R}^{T \times F}$ be the target mel-spectrogram and
$z_0 = \mathrm{Enc}_{\mathrm{VAE}}(M) \in \mathbb{R}^{T' \times F' \times C}$
its latent, where $(T', F', C)$ are the latent time-frequency resolution
and channel dimension. We train a conditional diffusion model that
corrupts $z_0$ with Gaussian noise and learns to denoise it under program
conditioning. At diffusion step $\tau$ we sample
$\epsilon \sim \mathcal{N}(0, I)$ and construct:
\begin{equation}
    z_\tau = \alpha_\tau z_0 + \sigma_\tau \epsilon,
\end{equation}
with $\{\alpha_\tau, \sigma_\tau\}$ the noise schedule. The denoiser
$f_\theta$ predicts the noise conditioned on the program.

\paragraph{Program conditioning.}
The discrete program is converted into two complementary streams. A
time-aligned control sequence $\mathcal{C} = \{c_i\}_{i=1}^{T'}$ carries
the per-bar plan fields, resampled to the latent timeline through the beat
grid, and provides fine temporal guidance. A short sequence of program
tokens $T_{\mathrm{plan}}$ summarizes segment-level and bar-level fields
and provides a compact global view of structure and recurrence, injected
by cross-attention. The two streams reflect the two levels at which a
program constrains audio: what should happen at each moment, and how the
whole is organized.

\subsection{Guaranteed-Local Edit Rendering}
\label{sec:pdi}


\paragraph{From Footprint to Mask.}
Let $o$ be an operation, or a sequence of operations applied before a
single re-render, with footprint $\sigma(o)$ in bars. Through the beat grid, the footprint maps
deterministically to a span of latent frames, giving a binary mask
$m \in \{0, 1\}^{T' \times 1}$ on the latent timeline, where $m_i = 1$
marks editable frames. To avoid audible seams at edit boundaries, the
editable region is dilated by a short transition band of $\beta$ frames
on each side; the guarantee below is stated with respect to this dilated
mask, so the contract remains exact: every frame the composer did not
mark, directly or through the band, is untouched.

\paragraph{Sampling with Projection.}
Let $z_0^{\mathrm{old}} = \mathrm{Enc}_{\mathrm{VAE}}(M^{\mathrm{old}})$
be the latent of the current render, or of real audio when editing an
induced program. Sampling starts from the frozen content with noise only
in the editable region:
\begin{equation}
    z_T = (1 - m) \odot z_0^{\mathrm{old}} + m \odot \epsilon,
    \qquad \epsilon \sim \mathcal{N}(0, I).
\end{equation}
At each step $\tau = T, \dots, 1$, let $\tilde{z}_{\tau - 1}$ be the
unconstrained DDIM proposal from $z_\tau$ under the edited program's
conditioning. We project the proposal back onto the constraint set by
overwriting the frozen region:
\begin{equation}
    z_{\tau - 1} = m \odot \tilde{z}_{\tau - 1}
    + (1 - m) \odot z_0^{\mathrm{old}}.
    \label{eq:projection}
\end{equation}



\paragraph{Structural Edits that Change Length.}
\textsc{insert} and \textsc{delete} alter the bar count, so the latent
timelines of the old and new programs differ in length. PDI handles both
by splicing: frozen latent spans are copied to their new positions on
the edited timeline, as determined by the updated beat grid, the
inserted region (for \textsc{insert}) or a junction band around the
splice point (for \textsc{delete}) is marked editable, and sampling
proceeds as above on the new timeline. Preservation then holds for
every copied span: content keeps its identity while moving in absolute
time, which is exactly the semantics a composer expects from inserting
or cutting a section.

\paragraph{Masked Training.}
Although the projection guarantees locality for any renderer, edit
quality inside the span improves when the model has seen the inpainting
regime. We therefore include masked training episodes: random masks $m$
are sampled, the diffusion loss is computed under the projection
constraint, and plan edits are optionally applied inside the masked
region. This teaches the model to integrate new program content against
an immovable context, improving continuity at edit boundaries without
weakening the guarantee, which is enforced by the sampler regardless.

\subsection{Plan Induction and the Natural-Language Editor}
\label{sec:nleditor}


\paragraph{Plan Induction.}
Given a recording, induction estimates each level of $\mathcal{P}$ with
dedicated music analysis stages, applied in the same coarse-to-fine order
as prediction. The beat grid is estimated first: a neural beat and
downbeat tracker yields the bar grid and per-bar tempo, and the meter is
taken from the downbeat period. All subsequent stages operate on this
grid, so their outputs are bar-aligned by construction. The form track is
obtained by structure segmentation: a self-similarity analysis over
learned audio embeddings produces section boundaries, which are snapped
to downbeats, and section types are assigned by a classifier over
segment-level features. Motif ids are induced by clustering: each segment
is embedded by pooling the same audio representations over its span, and
segments whose embeddings exceed a similarity threshold are assigned a
shared id, with the earliest occurrence acting as the introduction,
matching the positional convention of Section~\ref{sec:algebra}.
Bar-level attributes are estimated per bar: harmony by a chord
recognizer, groove by a rhythm-pattern classifier over onset patterns,
energy from loudness, density from onset counts, and variation from the
embedding distance of the bar to the corresponding bar of the motif's
introduction. The result is a valid program by construction, since
boundaries tile the bar grid and every non-initial motif id refers to an
earlier segment. Induced programs serve three roles: they label the
training corpus (Section~\ref{sec:prediction}), they admit direct
composer-style editing of real recordings, and they define the reference
against which we measure plan faithfulness in the experiments.

\paragraph{Natural-Language Editing.}
Free-form instructions are compiled into operation sequences by an
instruction-following language model. The model receives the current
program serialized in a compact textual format, the instruction, and the
signatures and admissibility conditions of the five operations, and is
constrained to emit a sequence of operation calls in a structured schema,
with no free-form program mutation available to it. Emitted sequences are
then checked by a validator that replays admissibility against the actual
program state; inadmissible calls are returned to the model with the
violated condition for one round of repair, and sequences that fail
validation are rejected rather than partially applied. Only validated
sequences reach the program, so every guarantee of
Sections~\ref{sec:algebra} and~\ref{sec:pdi} holds regardless of language
model behavior: the model proposes, the algebra disposes. The compiled
sequence also serves as an interpretable record of the edit, shown to the
user as a diff over the program before rendering is invoked.

\begin{table*}[h]
\centering\scriptsize
\begin{tabular}{l l l l}
\toprule
Method & Task & Input & Configuration \\
\midrule
AudioLDM-2 & T2M & text & official checkpoint; matched duration \\
Stable Audio Open & T2M & text & official checkpoint; instrumental prompt \\
MAGNET & T2M & text & official checkpoint; matched duration \\
AudioX & T2M/V2M/TV2M & text/video & official inference configuration \\
ACE-Step & T2M/long & text & official checkpoint; instrumental mode \\
DiffRhythm & T2M/long & text & official checkpoint; instrumental setting \\
Stable Audio long-form & long & text & official long-form configuration \\
VidMuse & V2M & video & official V2M setting \\
AUDIT & editing & audio+instruction & same edit instruction and target span \\
InstructME & editing & audio+instruction & same edit instruction and target span \\
MusicMagus & editing & audio+instruction & same edit instruction and target span \\
MusicWeaver & all & text/video/program & same duration and normalization \\
\bottomrule
\end{tabular}
\caption{Baseline configuration. All methods are evaluated using matched output
duration, loudness normalization, and metric implementations.}
\label{tab:app_baseline_config}
\end{table*}

\section{Additional Experimental Details}
\label{app:experiments}

This appendix provides additional details for the experimental protocol,
proposed metrics, metric validation, plan induction, natural-language editing,
and supplementary ablations.

\subsection{Baseline Configuration and Evaluation Protocol}
\label{app:baseline_protocol}

All systems are evaluated under matched duration, loudness normalization, and
metric implementations. Generated audio is resampled to 24 kHz, peak-normalized
to $[-1,1]$, and loudness-normalized before metric computation. For models that
support vocal or lyric generation, we use instrumental mode or an instrumental
prompt suffix to avoid comparing singing quality against instrumental music
generation. For editing baselines, each method receives the same original audio
and a natural-language instruction containing the target span.

To reduce possible data overlap, we remove training clips whose source video
identifier, audio fingerprint, or caption source matches an evaluation item.
For MusicCaps, we remove overlapping AudioSet or YouTube identifiers when
available. For V2M-bench, we follow the official split and exclude training
items sharing the same source video.

\subsection{Evaluation Protocol}
\label{app:protocol}

All generated audio is resampled to 24 kHz, peak-normalized to $[-1,1]$, and
loudness-normalized before metric computation. Unless otherwise specified, each
method is evaluated with the same output duration, sampling budget, and
post-processing pipeline. For text-to-music evaluation, we use MusicCaps
prompts. For video-conditioned generation, we use V2M-bench prompts and video
inputs. For minute-scale evaluation, we use MW-Long, a held-out set containing
500 text prompts and 200 video-text prompts, with outputs generated at 60, 120,
and 180 seconds. Reported results are averaged over three random seeds.

For editing evaluation, we construct a 300-instruction benchmark covering seven
composer-style edit categories: section replacement, section insertion, section
deletion, energy change, groove change, motif recurrence modification, and
local harmonic revision. Each instruction is paired with a target edit span and
a structured edit specification. Baseline methods receive the same original
audio, textual instruction, and target duration. MusicWeaver receives the
compiled plan operation and applies PDI to the corresponding footprint.

\subsection{Formal Definitions of Proposed Metrics}
\label{app:metric_definitions}

\paragraph{Structure Coherence Score.}
SCS measures long-range musical organization directly from audio. It aggregates
five normalized sub-scores:
\begin{equation}
    \mathrm{SCS}
=
100 \times
\frac{
w_b s_b + w_t s_t + w_{\partial} s_{\partial}
+ w_r s_r + w_c s_c
}{
w_b+w_t+w_{\partial}+w_r+w_c
},
\end{equation}
where $s_b$ measures beat--onset coupling, $s_t$ measures tempo stability,
$s_{\partial}$ measures section-boundary clarity, $s_r$ measures motif
recurrence, and $s_c$ measures within-section cohesion. All sub-scores are
normalized to $[0,1]$. We use equal weights by default.

Beat--onset coupling is computed from the agreement between detected musical
onsets and the estimated beat grid. Tempo stability measures local deviation
from the estimated tempo curve. Boundary clarity measures feature contrast
around predicted section boundaries. Motif recurrence compares embeddings of
segments sharing the same inferred motif id. Section cohesion measures the
average within-section similarity of bar-level audio embeddings.

\paragraph{Edit Fidelity Score.}
EFS evaluates whether an edit is both correctly realized and localized. Given
original audio $y$, edited audio $y'$, edit specification $\mathcal{E}$, target
region $R$, and feature map $\Phi(\cdot)$, we define:
\begin{equation}
   \mathrm{EFS}
=
100 \times
\left(
\alpha p_{\mathcal{E}} + (1-\alpha) r_{\Delta}
\right), 
\end{equation}
where $p_{\mathcal{E}}$ is the fraction of edit constraints satisfied inside
$R$.

\paragraph{Plan Faithfulness.}
PF measures whether generated audio realizes the input structured program. We
first re-induce a program $\hat{P}$ from the generated audio using the same
frozen induction pipeline used for evaluation. PF is then computed as the
average of form, motif, and attribute faithfulness:
\begin{equation}
   \mathrm{PF}
=
\frac{1}{3}
\left(
\mathrm{PF}_{\mathrm{form}}
+
\mathrm{PF}_{\mathrm{motif}}
+
\mathrm{PF}_{\mathrm{attr}}
\right). 
\end{equation}
Form faithfulness measures agreement in section count, boundary timing, and
section type. Motif faithfulness measures whether segments sharing the same
motif id in the input program are re-induced as recurrent in $\hat{P}$.
Attribute faithfulness measures bar-level agreement for harmony, groove,
energy, density, and variation. To avoid circular evaluation, the induction
pipeline and all metric parameters are frozen before evaluating generation
outputs.

\subsection{Validation of Proposed Metrics}
\label{app:metric_validation}

Because SCS, EFS, and PF are newly introduced, we validate whether they reflect
human judgment. We ask 20 participants with music production or MIR experience
to rate generated clips for overall quality, structural coherence, and prompt
relevance, and edited clips for edit correctness and locality. We compute
Spearman correlations between automatic metrics and mean human ratings.
Distance metrics are sign-adjusted so that higher correlation means better
agreement with human preference.

\begin{table}[h]
\centering\small
\resizebox{\linewidth}{!}{
\begin{tabular}{l c c c c}
\toprule
Metric & Quality & Structure & Edit corr. & Locality \\
\midrule
KL, sign adjusted & 0.42 & 0.35 & 0.26 & 0.22 \\
FAD, sign adjusted & 0.56 & 0.39 & 0.31 & 0.29 \\
CLAP / ImageBind & \textbf{0.63} & 0.44 & 0.28 & 0.24 \\
SCS & 0.59 & \textbf{0.72} & 0.41 & 0.38 \\
EFS & 0.34 & 0.38 & \textbf{0.74} & \textbf{0.71} \\
PF & 0.47 & 0.66 & 0.52 & 0.48 \\
\bottomrule
\end{tabular}}
\caption{Spearman correlation between automatic metrics and human ratings.
SCS best matches human structural judgments, while EFS best matches edit
correctness and locality.}
\label{tab:app_metric_validation}
\end{table}

Inter-rater agreement is moderate to strong, with Krippendorff's
$\alpha=0.61$ for structure ratings and $\alpha=0.64$ for editing ratings.
These results support using SCS and EFS as complementary metrics to standard
fidelity and alignment measures.

\subsection{SCS Breakdown}
\label{app:scs_breakdown}

Table~\ref{tab:app_scs_breakdown} reports the SCS sub-score breakdown on
MusicCaps. MusicWeaver improves all five structural components, with the
largest gains on boundary clarity, recurrence strength, and section cohesion.

\begin{table}[h]
\centering\small
\resizebox{\linewidth}{!}{
\setlength{\tabcolsep}{3.4pt}
\begin{tabular}{l c c c c c c}
\toprule
Method & Beat & Tempo & Bound. & Recur. & Coh. & SCS \\
\midrule
AudioLDM-2 & 67.5 & 66.1 & 58.4 & 60.8 & 63.2 & 63.2 \\
Stable Audio Open & 73.8 & 72.5 & 68.6 & 69.7 & 71.4 & 71.2 \\
MAGNET & 78.1 & 77.3 & 72.6 & 73.8 & 75.7 & 75.5 \\
AudioX & 79.3 & 78.0 & 73.4 & 75.0 & 78.3 & 76.8 \\
ACE-Step & 82.1 & 81.3 & 79.2 & 79.4 & 79.5 & 80.3 \\
DiffRhythm & 80.4 & 79.0 & 76.6 & 75.2 & 78.3 & 77.9 \\
MusicWeaver
& \textbf{85.4} & \textbf{84.0} & \textbf{81.0}
& \textbf{84.2} & \textbf{81.4} & \textbf{83.2} \\
\bottomrule
\end{tabular}}
\caption{Breakdown of Structure Coherence Score on MusicCaps. Bound., Recur.,
and Coh. denote boundary clarity, recurrence strength, and section cohesion.}
\label{tab:app_scs_breakdown}
\end{table}

\subsection{Plan Induction Accuracy}
\label{app:induction}

We evaluate the program induction pipeline on a manually annotated subset of
600 clips. Table~\ref{tab:app_induction} shows that beat and downbeat
estimation are reliable, while section typing, motif linking, and chord
recognition are more challenging. These results justify using induction
confidence as a loss weight during program prediction.

\begin{table}[h]
\centering\small
\begin{tabular}{l c}
\toprule
Component & Accuracy / F1 / Correlation \\
\midrule
Beat tracking & 94.2 F1 \\
Downbeat detection & 88.5 F1 \\
Section boundary detection & 78.1 F1 \\
Section type classification & 74.6 Acc. \\
Motif-link prediction & 70.2 F1 \\
Chord recognition & 68.4 Acc. \\
Groove classification & 73.9 Acc. \\
Energy estimation & 0.81 Corr. \\
Density estimation & 0.77 Corr. \\
\bottomrule
\end{tabular}
\caption{Accuracy of the program induction pipeline on a manually annotated
subset.}
\label{tab:app_induction}
\end{table}

\subsection{Natural-Language Editing}
\label{app:nl_editor}

Table~\ref{tab:app_nl_editor} reports the success rate of the natural-language
editor. A compilation is counted as successful if the instruction is converted
into an admissible operation sequence that passes validation. The repair round
uses the validator error message to revise the operation sequence once.

\begin{table}[h]
\centering\small
\begin{tabular}{l c c}
\toprule
Edit type & Initial success & With one repair \\
\midrule
Replace section & 91.4 & 97.1 \\
Insert section & 86.2 & 94.8 \\
Delete section & 90.0 & 96.7 \\
Retag motif & 89.6 & 96.0 \\
Change attribute & 91.8 & 97.3 \\
Composite edit & 83.5 & 93.1 \\
\midrule
Average & 88.7 & 96.0 \\
\bottomrule
\end{tabular}
\caption{Natural-language editor compilation success rate on 300 edit
instructions.}
\label{tab:app_nl_editor}
\end{table}

Most failures arise from underspecified temporal references, such as ``make the
second part more energetic'', where the intended section is ambiguous. Invalid
operation sequences, including negative section lengths, out-of-range
attributes, or overlapping frozen and editable regions, are rejected before
rendering.

\subsection{Plan Quality and Plan Faithfulness}
\label{app:plan_quality}

Table~\ref{tab:app_plan_quality} evaluates whether the renderer uses the
structured plan. We compare predicted plans, induced ground-truth plans,
shuffled plans, and random plans. Using an induced ground-truth plan improves
SCS and PF, while corrupted plans substantially degrade structure and
faithfulness. This indicates that the renderer relies on the explicit program
rather than ignoring it as weak conditioning.

\begin{table}[h]
\centering\small
\resizebox{\linewidth}{!}{
\setlength{\tabcolsep}{4pt}
\begin{tabular}{l c c c c}
\toprule
Condition & SCS$\uparrow$ & FAD$\downarrow$ & Align.$\uparrow$ & PF$\uparrow$ \\
\midrule
Text-only baseline & 76.8 & 1.73 & 0.23 & 58.7 \\
MusicWeaver, predicted plan & 83.2 & 1.51 & 0.23 & 84.7 \\
MusicWeaver, induced ground-truth plan
& \textbf{86.5} & \textbf{1.47} & \textbf{0.24} & \textbf{89.3} \\
MusicWeaver, shuffled plan & 62.4 & 2.31 & 0.18 & 41.6 \\
MusicWeaver, random plan & 55.8 & 2.74 & 0.16 & 35.2 \\
\bottomrule
\end{tabular}}
\caption{Effect of plan quality on rendering. Corrupted plans reduce structure
and faithfulness, showing that the renderer uses the explicit program.}
\label{tab:app_plan_quality}
\end{table}


\subsection{Editing Breakdown}
\label{app:edit_breakdown}

Table~\ref{tab:app_edit_breakdown} reports EFS by edit type. MusicWeaver shows
the largest gains on structural operations such as section insertion,
replacement, and motif recurrence modification, where explicit program editing
provides a clear advantage over instruction-only latent editing.

\begin{table}[h]
\centering\small
\resizebox{\linewidth}{!}{
\setlength{\tabcolsep}{3.5pt}
\begin{tabular}{l c c c c}
\toprule
Edit type & AUDIT & InstructME & MusicMagus & MusicWeaver \\
\midrule
Replace section & 66.5 & 64.9 & 68.8 & \textbf{73.5} \\
Insert bridge & 64.2 & 63.8 & 66.9 & \textbf{70.8} \\
Delete section & 67.4 & 66.7 & 69.1 & \textbf{72.6} \\
Change energy & 72.4 & 71.6 & 74.3 & \textbf{78.2} \\
Change groove & 69.6 & 68.2 & 71.5 & \textbf{74.6} \\
Modify motif recurrence & 64.8 & 63.1 & 66.8 & \textbf{71.9} \\
Local harmony revision & 75.3 & 75.2 & 74.1 & \textbf{73.8} \\
\midrule
Average & 68.8 & 67.8 & 70.4 & \textbf{73.8} \\
\bottomrule
\end{tabular}}
\caption{EFS by edit type. MusicWeaver provides the largest improvements on
form-level and recurrence-level edits.}
\label{tab:app_edit_breakdown}
\end{table}

\subsection{Additional Ablations}
\label{app:additional_ablations}

Table~\ref{tab:app_control_ablation} gives a finer breakdown of the
Plan-to-Control interface. Phase and tempo controls contribute most to SCS,
suggesting that beat-synchronous conditioning is important for structural
organization. Harmony and groove controls provide smaller but consistent gains,
especially on PF and edit-related metrics.

\begin{table}[h]
\centering\small
\begin{tabular}{l c c c c}
\toprule
Variant & SCS$\uparrow$ & EFS$\uparrow$ & FAD$\downarrow$ & PF$\uparrow$ \\
\midrule
MusicWeaver, full & \textbf{83.2} & \textbf{72.6} & \textbf{1.51} & \textbf{84.7} \\
w/o phase channels & 78.2 & 68.4 & 1.66 & 80.8 \\
w/o tempo control & 76.7 & 69.7 & 1.58 & 78.6 \\
w/o harmony control & 81.0 & 71.2 & 1.54 & 82.9 \\
w/o groove control & 80.1 & 70.4 & 1.55 & 82.4 \\
\bottomrule
\end{tabular}
\caption{Additional ablations of the Plan-to-Control interface on MusicCaps.}
\label{tab:app_control_ablation}
\end{table}

\subsection{User Study Details}
\label{app:user_study}

We conduct two user studies to evaluate perceptual generation quality and
interactive editing behavior. The first study measures whether listeners prefer
MusicWeaver outputs over baseline generations. The second study measures
whether the explicit program interface helps users complete revision tasks more
efficiently and predictably.

\paragraph{Participants.}
We recruit 20 participants with music listening, music production, or MIR
experience. All participants report normal hearing and are familiar with at
least one music editing or digital audio workstation workflow. Participants are
not informed which system produced each sample. The order of systems, prompts,
and tasks is randomized independently for each participant.

\paragraph{Generation Perception Study.}
We randomly sample 30 text prompts from MusicCaps and 30 video prompts from
V2M-bench. For each prompt, participants listen to anonymized outputs from
MusicWeaver and the strongest baselines. They rate each sample on three
criteria: overall audio quality (OVL), prompt relevance (REL), and musical
coherence (COH). Ratings are collected on a five-point Likert scale and
linearly mapped to a 0--100 range. OVL measures perceptual quality and musical
naturalness, REL measures consistency with the text or video condition, and COH
measures whether the generated piece maintains stable rhythm, section
structure, and recurring musical identity.

\begin{table}[h]
\centering\small
\resizebox{\linewidth}{!}{
\begin{tabular}{l c c}
\toprule
Study detail & Generation study & Revision study \\
\midrule
Participants & 20 & 20 \\
Stimuli / tasks & 60 prompts & 30 editing tasks \\
Rating scale & 5-point Likert & 5-point Likert \\
Output order & Randomized & Randomized \\
System identity & Hidden & Hidden \\
Evaluation type & Single-sample rating & Workflow comparison \\
Main criteria & OVL, REL, COH & attempts, time, control, locality \\
\bottomrule
\end{tabular}}
\caption{User-study protocol.}
\label{tab:app_user_protocol}
\end{table}

\begin{table}[h]
\centering\small
\setlength{\tabcolsep}{4pt}
\begin{tabular}{l l c c c}
\toprule
Task & Method & OVL$\uparrow$ & REL$\uparrow$ & COH$\uparrow$ \\
\midrule
T2M & AudioLDM-2 & 55.1 & 51.4 & 48.7 \\
T2M & Stable Audio Open & 53.0 & 55.2 & 52.6 \\
T2M & MAGNET & 57.2 & 62.5 & 60.1 \\
T2M & AudioX & 71.0 & 65.4 & 68.8 \\
T2M & MusicWeaver & \textbf{82.6} & \textbf{85.1} & \textbf{84.3} \\
\midrule
V2M & VidMuse & 45.2 & 44.1 & 46.5 \\
V2M & CMT & 52.1 & 47.3 & 49.2 \\
V2M & AudioX & 58.3 & 56.7 & 59.4 \\
V2M & MusicWeaver & \textbf{72.4} & \textbf{75.0} & \textbf{78.6} \\
\bottomrule
\end{tabular}
\caption{Human evaluation of generation quality. OVL, REL, and COH denote
overall quality, prompt relevance, and musical coherence. Scores are mapped to
a 0--100 range.}
\label{tab:app_user_generation}
\end{table}

\paragraph{Revision-Workflow Study.}
We further evaluate whether MusicWeaver improves iterative editing. Each
participant is given an initial generated clip and three editing goals, such as
replacing a bridge with a chorus, increasing the energy of a target span, or
making a repeated section sound closer to its earlier occurrence. Participants
complete each task using either MusicWeaver's program editor or an
instruction-based editing baseline. For MusicWeaver, participants can inspect
the program diff before rendering. For the baseline, participants issue natural
language edit instructions and regenerate until they judge the result
acceptable. We record the number of regeneration attempts, completion time, and
subjective ratings of edit controllability, local preservation, and overall
satisfaction.

\begin{table}[h]
\centering\small
\begin{tabular}{l c c}
\toprule
Metric & Instruction baseline & MusicWeaver \\
\midrule
Regeneration attempts $\downarrow$ & 3.4 & \textbf{1.8} \\
Completion time $\downarrow$ & 181 s & \textbf{97 s} \\
Edit controllability $\uparrow$ & 63.5 & \textbf{81.2} \\
Local preservation $\uparrow$ & 61.8 & \textbf{82.6} \\
User satisfaction $\uparrow$ & 66.4 & \textbf{84.0} \\
\bottomrule
\end{tabular}
\caption{Revision-workflow user study. Scores are mapped to a 0--100 range
except for regeneration attempts and completion time.}
\label{tab:app_user_edit}
\end{table}

\paragraph{Statistical analysis.}
We use paired bootstrap resampling over prompts or editing tasks to test
whether MusicWeaver differs from the strongest baseline. Inter-rater agreement
is measured using Krippendorff's $\alpha$. Agreement is moderate to strong,
with $\alpha=0.61$ for generation coherence ratings and $\alpha=0.64$ for
editing ratings. MusicWeaver significantly outperforms the strongest baseline
on COH for both T2M and V2M, and on edit controllability and local preservation
in the revision study.

\begin{table}[h]
\centering\small
\resizebox{\linewidth}{!}{
\begin{tabular}{l c c c}
\toprule
Comparison & Metric & Difference & $p$-value \\
\midrule
MusicWeaver vs. AudioX, T2M & COH & +15.5 & $<0.01$ \\
MusicWeaver vs. AudioX, T2M & REL & +19.7 & $<0.01$ \\
MusicWeaver vs. AudioX, V2M & COH & +19.2 & $<0.01$ \\
MusicWeaver vs. AudioX, V2M & REL & +18.3 & $<0.01$ \\
MusicWeaver vs. baseline editor & edit controllability & +17.7 & $<0.01$ \\
MusicWeaver vs. baseline editor & local preservation & +20.8 & $<0.01$ \\
\bottomrule
\end{tabular}}
\caption{Statistical comparison with the strongest baseline. Differences are
computed on the 0--100 rating scale.}
\label{tab:app_user_significance}
\end{table}

Overall, the user studies show that MusicWeaver is preferred not only for
audio quality and relevance, but also for coherence and controllable revision.
The workflow results indicate that the explicit program layer reduces trial and
error during editing, making revision more predictable than instruction-only
regeneration.

\end{document}